\documentclass[12pt,a4paper]{article}
\pdfoutput=1

\usepackage{amsmath,amssymb}
\usepackage[bookmarks=true]{hyperref}
\usepackage[nosort]{cite}
\usepackage[bulletsep]{collect}
\def\gfxon{\usepackage[final]{graphicx}}

\gfxon

\makeatletter
\let\old@startsection=\@startsection
\renewcommand{\@startsection}[6]{\old@startsection{#1}{#2}{#3}{#4}{#5}{#6\mathversion{bold}}}
\makeatother

\newcommand{\po}[1]{\frac{\partial}{\partial #1}}

\newcommand{\dpou}[1]{\partial^{#1}}
\newcommand{\dpod}[1]{\partial_{#1}}

\newcommand{\eps}{\varepsilon}

\makeatletter \@addtoreset{equation}{section} \makeatother

\makeatletter
\let\old@makecaption=\@makecaption
\def\@makecaption{\small\old@makecaption}
\makeatother

\newcommand{\ham}{\mathcal{H}}

\newcommand{\gen}[1]{\mathfrak{#1}}

\newcommand{\moment}{\mathcal{P}}

\newcommand{\Bfield}{\mathcal{B}}

\newcommand{\vecs}{\mathcal{N}}
\newcommand{\const}{\mathcal{S}}

\newcommand{\energy}{\mathcal{E}}

\makeatletter
\newcommand{\ellSN}{\mathop{\operator@font sn}\nolimits}
\newcommand{\ellCN}{\mathop{\operator@font cn}\nolimits}
\newcommand{\ellDN}{\mathop{\operator@font dn}\nolimits}
\newcommand{\ellAM}{\mathop{\operator@font am}\nolimits}
\newcommand{\ellK}{\mathop{\smash{\operator@font K}\vphantom{a}}\nolimits}
\newcommand{\ellE}{\mathop{\smash{\operator@font E}\vphantom{a}}\nolimits}
\makeatother

\ifx\genfrac\sdflkaj

\else

\fi
\newcommand{\sfrac}[2]{{\textstyle\frac{#1}{#2}}}
\newcommand{\half}{\sfrac{1}{2}}

\newcommand{\quarter}{\sfrac{1}{4}}

\newcommand{\acomm}[2]{\{#1,#2\}}

\newcommand{\covder}{\mathcal{D}}

\newcommand{\nln}{\nonumber\\}
\newcommand{\nl}[1][0pt]{\nonumber\\[#1]&\hspace{-4\arraycolsep}&\mathord{}}

\newcommand{\earel}[1]{\mathrel{}&\hspace{-2\arraycolsep}#1\hspace{-2\arraycolsep}&\mathrel{}}
\newcommand{\eq}{\earel{=}}
\newcommand{\beq}{\begin{equation}}
\newcommand{\eeq}{\end{equation}}

\def\[{\begin{equation}}
\def\]{\end{equation}}
\def\<{\begin{eqnarray}}
\def\>{\end{eqnarray}}

\def\a{\alpha}
\def\b{\beta}

\def\eps{\varepsilon}
\def\g{\gamma}

\def\m{\mu}
\def\p{\partial}
\def\s{\sigma}
\def\t{\theta}
\def\z{\zeta}

\def\f{\frac}
\def\be{\begin{equation}}
\def\ee{\end{equation}}
\def\bea{\begin{eqnarray}}
\def\eea{\end{eqnarray}}
\def\ba{\begin{array}}
\def\ea{\end{array}}
\def\bc{\begin{center}}
\def\ec{\end{center}}
\def\bl{\begin{flushleft}}
\def\el{\end{flushleft}}
\def\br{\begin{flushright}}
\def\er{\end{flushright}}
\def\l{\left}
\def\r{\right}

\makeatletter
\def\mr@ignsp#1 {\ifx\:#1\@empty\else #1\expandafter\mr@ignsp\fi}%
\newcommand{\multiref}[1]{\begingroup
\xdef\mr@no@sparg{\expandafter\mr@ignsp#1 \: }%
\def\mr@comma{}%
\@for\mr@refs:=\mr@no@sparg\do{\mr@comma\def\mr@comma{,}\ref{\mr@refs}}%
\endgroup}
\makeatother

\newcommand{\secref}[1]{Sec.~\multiref{#1}}
\newcommand{\Appref}[1]{Appendix~\multiref{#1}}

\newcommand{\figref}[1]{Fig.~\multiref{#1}}
\renewcommand{\eqref}[1]{(\multiref{#1})}
\newcommand{\lagr}{\mathcal{L}}

\ifx\href\asklfhas\newcommand{\href}[2]{#2}\fi
\newcommand{\arxivno}[1]{\href{http://arxiv.org/abs/#1}{#1}}

\begin{document}

\begin{flushright}\footnotesize
\texttt{ArXiv:\arxivno{1003.2645}}\\
\texttt{FTPI-MINN-10/08} \\
\texttt{UMN-TH-2839/10}
\vspace{0.5cm}
\end{flushright}
\vspace{0.3cm}

\renewcommand{\thefootnote}{\arabic{footnote}}
\setcounter{footnote}{0}
\begin{center}%
{\Large\textbf{\mathversion{bold}
Large-N Solution of the Heterotic $\mathcal{N}=(0,1)$ Two-Dimensional O(N) Sigma Model}
\par}

\vspace{1cm}%

\textsc{Peter Koroteev and Alexander Monin}

\vspace{5mm}

\textit{University of Minnesota, School of Physics and Astronomy\\%
116 Church Street S.E. Minneapolis, MN 55455, USA}

\vspace{7mm}

\thispagestyle{empty}

\texttt{koroteev@physics.umn.edu} \\ 
\texttt{monin@physics.umn.edu}

\par\vspace{1cm}

\vfill

\textbf{Abstract}\vspace{5mm}

\begin{minipage}{12.7cm}
In this paper we build a family of heterotic deformations of the $\text{O}(N)$ sigma model.  These deformations break $(1,1)$ supersymmetry down to $(0,1)$ symmetry. We solve this model at large $N$. We also find an alternative superfield formulation of the heterotic $\mathbb{CP}^{N}$ sigma model which was discussed in the literature before.  
\end{minipage}

\vspace{3mm}

\vspace*{\fill}

\end{center}

\newpage

\section{Introduction}\label{Sec:Intro}


Various nonlinear sigma models have being investigated due to their similarities with non-Abelian gauge theories. Polyakov introduced the $\text{O}(N)$ bosonic sigma model and showed that it is asymptotically free \cite{Polyakov:1975rr}. This model was solved exactly for $N=3$ in Zamolodchikov's paper \cite{Zamolodchikov:1978xm}. The supersymmetric generalization of the $\text{O}(3)$ sigma model was constructed by Witten in \cite{Witten:1977xn}. He also developed a technique for solving the $\mathbb{CP}^{N}$ sigma model at large $N$ for both supersymmetric and non-supersymmetric theories \cite{Witten:1978bc}. The common feature of the sigma models (without a mass), apart from being asymptotically free, is that due to quantum effects the spontaneous symmetry breaking disappears, and the symmetry gets restored. This fact is reflected by the absence of massless (not sterile) particles in the theory.

Two-dimensional sigma-models with $(0,2)$ supersymmetry have been considered some time ago \cite{Witten:1993yc, Hull:1985jv}\footnote{More references about study of superspaces, renormalization of heterotic sigma models can be found in \cite{West:1986wua}}. More recently, the new interest in the heterotic $\mathbb{CP}^{N}$ sigma model was induced by the fact that this model naturally emerges as an effective theory of the moduli on the world-sheet of the non-Abelian flux tubes \cite{Gorsky:2004ad}  which are present in four dimensional $\text{SU}(N)$ Yang-Mills theories \cite{Hanany:2003hp,Hanany:2004ea,Shifman:2004dr,Auzzi:2003fs,Eto:2005yh}(for a review see \cite{Shifman:2007ce, Tong:2008qd,Eto:2006pg}). The non-Abelian vortices for arbitrary gauge group in particular for $\text{SO}(N)$ and $\text{USp}(N)$ were constructed in \cite{Eto:2009bg,Eto:2008yi}, while the generalization for the higher winding numbers was done in \cite{Eto:2006cx}.  Edalati and Tong suggested \cite{Edalati:2007vk} that the world sheet theory is a heterotic $\mathcal{N} = (0,2)$ theory. It was proven to be correct in the papers \cite{Shifman:2008wv, Bolokhov:2009sg}, where the heterotic model was explicitly obtained from the $\mathcal{N}=2$ SYM bulk theory deformed by the mass terms of the adjoint fields. Since then the different aspects of the $\mathbb{CP}^{N}$ sigma model were considered in great details \cite{Gorsky:2005ac,Shifman:2008kj,Bolokhov:2009wv,Bolokhov:2010hv,Shifman:2010id}.

In the present work, inspired by the recent developments in heterotic $\mathbb{CP}^{N}$ models, we study heterotic deformation of a supersymmetric $\text{O}(N)$ sigma model. The undeformed theory contains $N$ scalar and $N$ spinor real-valued fields. The bosonic fields are confined on a $(N-1)$-sphere. For generic values of $N$ only $(1,1)$ supersymmetry is present in the model, while for $N=3$ the model is equivalent to the supersymmetric $\mathbb{CP}^1$ sigma model, which actually possesses $\mathcal {N} = (2,2)$ supersymmetry\footnote{Due to the K\"{a}hler structure of the target manifold, which is $S ^ 2$ in this case, the extra supercurrent emerges lifting $(1,1)$ supersymmetry up to $(2,2)$.}. 
Although, to our knowledge, no bulk theory has been constructed for the heterotic $\text{O}(N)$ supersymmetric sigma model \footnote{However, there is an example \cite{Markov:2004mj} when the $\text{O}(3)$ sigma model emerges as an effective theory on the world-sheet of the string in the $\mathcal{N}=1^*$ supersymmetric $\text{SU}(2)$ gauge theory. But as we know the $N=3$ case is special.}-- it is interesting in its own right.

Given these obvious differences between $\text{SU}(N)$ and $\text{O}(N)$ models, one may expect to see different physical properties in the large N limit. In the current paper we find the spectrum of both $(1,1)$ and $(0,1)$ supersymmetric  $\text{O}(N)$ sigma models and observe that it is very much reminiscent of the $\text{SU}(N)$ models. The only major difference between the two models is the number of vacua -- it is always two in the orthogonal case and $N$ in the unitary case (for the $\mathbb{CP}^{N-1}$ model). 

 
The paper is organized as follows.  In \secref{Sec:ONsigmamodel} we discuss the $(1,1)$ supersymmetric $\text{O}(N)$ sigma model and introduce its heterotic deformation first in terms of superfields, then in components. In \secref{Sec:EffPotVac} we find the vacua of $(0,1)$ heterotic $\text{O}(N)$ model and derive the effective potential. \secref{Sec:Spectrum} is devoted to the investigation of the spectrum of the model in question, we give explicit formulas for the masses and the couplings at different values of the deformation parameter. In \secref{Sec:SuperfieldCPN} both the (0,2) heterotic $\mathbb{CP}^{N-1}$ and the weighted $\mathbb{CP}^{N-1}$ sigma models, which were discussed in \cite{Shifman:2008kj} and \cite{Shifman:2010id}, are formulated using the superfields. In \secref{Sec:Conclusions} we present conclusions, and \Appref{Sec:Notations} contains notations and conventions we use in the paper.

\section{Supersymmetric $\text{O}(N)$ sigma model and its heterotic deformation}\label{Sec:ONsigmamodel}

The bosonic $\text{O}(N)$ sigma model with coupling constant $g_0$ can be formulated as follows. The dynamics of $N$ real-valued scalar fields, subject to the constraint 
\[
(n^i)^2 = 1,
\label{bcons}
\]
is governed by the action
\[\label{bact}
S = \frac{1}{4g_0^2}\int d^2 x\,\dpod{\mu}n^i \dpou{\mu}n ^ i\,.
\]
The constraint (\ref{bcons}) means that the isovector field $n^i\,,i=\overline{1,N}$ is confined on a unit $(N-1)$-sphere. The coupling constant $g_0$ in (\ref{bact}) is a bare one. Considering additional fermionic degrees of freedom one can easily supersymmetrize the model \cite{Witten:1977xn}. Supersymmetry is obvious when the action is written in terms of the superfields \footnote{Our notations can be found in \Appref{Sec:Notations}.}
\bea
\label{eq:LagrNonChiral0}
\lagr \eq \frac{1}{4} \int d^2 \theta \, \varepsilon _ {\a \b} \covder _\beta \vecs^i \covder _ \alpha \vecs_i = 
2 \p _ L n ^ i \p _ R n ^ i + i \psi _ L ^ i \p _ R \psi _ L ^ i
+ i \psi _ R ^ i \p _ L \psi _ R ^ i + \frac{1}{2}(F^i)^2,
\eea
where the the so-called isovector superfield has the following components
\[\label{eq:IsovectorSuperField}
\vecs^i = n^i + \bar{\theta}\psi^i + \half \bar{\theta}\theta F^i\,,\qquad i=1,\dots ,N\,,
\]
with a generalization of the bosonic constraint (\ref{bcons})
\be
\vecs ^ 2 = r _ 0,
\label{sscons}
\ee
where a new coupling was introduced $r _ 0 = g _ 0 ^ {-2}$. In \eqref{eq:IsovectorSuperField} $\psi$ is a Majorana two-component spinor together with $\theta$ in \eqref{eq:LagrNonChiral0}. All components of the isovector superfield $N^i$ \eqref{eq:IsovectorSuperField} are real-valued.  We have rescaled the fields in such a way that the coupling constant appears in the constraint (\ref{sscons}) rather than in the action (\ref{eq:LagrNonChiral0}). Taking into account that
\be
\vecs ^ 2 = n ^ i n ^ i + 2 \bar \theta \psi ^ i n ^ i + 
\bar \t \t \l (F ^ i n^ i - \half \bar \psi ^ i \psi ^ i \r ),
\ee
one can write the relations for the components of the superfields $\mathcal{N} ^ i$ in the following form
\bea
\label{eq:SphereConstr}
n^2 \eq r _ 0 \,,\nln
n^i\psi^\alpha_i \eq 0\,,\nln
F ^ i n^ i - \half\bar \psi ^ i \psi ^ i \eq 0.
\eea
The usual way to take into account the constraint is to introduce a Lagrange multiplier. For the case at hand the latter is a chiral superfield
\be
\mathcal{S} = \s + \bar \t \lambda + \half \bar \t \t D.
\ee
Again, as in \eqref{eq:IsovectorSuperField} all the components of the superfield $\const$ are real-valued.
Therefore the action (\ref {eq:LagrNonChiral0})can be rewritten in the following form
\be
\mathcal {L} = \int d^2 \theta\, \l [ \frac{1}{4} \, \varepsilon _ {\a \b} \covder _\b \vecs^i \covder _ \alpha \vecs_i 
+ \f {1} {4 e _ 0 ^ 2} \, \varepsilon _ {\a \b} \covder _ \b \mathcal{S}\, \covder_\alpha \mathcal {S}+ \frac{i}{2} \mathcal {S} \l ( \mathcal{N}^2 - r _ 0 \r ) \r],
\label{eq:LagrNonChiral}
\ee
where the limit for the coupling constant $e _ 0 ^ 2 \to \infty$ is implied. Hence the auxiliary fields are not dynamical for the time being, however, in \secref{Sec:Spectrum} it will be shown that the coupling constant gets renormalized, therefore, providing non vanishing kinetic terms for the auxiliary fields. Those will be used in investigating the mass spectra of the theory.

\paragraph{Heteroric deformation.}

The model \eqref{eq:LagrNonChiral} is $\mathcal{N}=(1,1)$ supersymmetric, namely it is invariant under both left-handed and right-handed transformations. Now we are going to deform it by adding an extra left-handed fermion mixing with the initial ones, obviously breaking the $(1,1)$ supersymmetry down to $(0,1)$. Using the language of the superfields, we add the new term which contains only left-handed fermion and an auxiliary field 
\[\label{eq:HeteroticDeformation}
\Delta \lagr = \int d^2\theta\, \left[ \quarter \, \varepsilon _ {\a \b} \covder _ \b \Bfield\, \covder _ \alpha \Bfield 
-i \gamma \const \Bfield \right]\,,
\]
where the chiral superfield $\mathcal {B}$ has the form
\[\label{eq:Bfield}
\Bfield = \bar \t \zeta + \half \bar \t \t G\,, \quad \zeta = \begin{pmatrix} 0 \\ \zeta_L \end{pmatrix}\,.
\]
It can be checked by a direct calculation that the above expression is indeed a superfield only with respect to the following left-handed transformations
\[
\delta \zeta_L = \varepsilon_L G\,,\quad \delta G = -2i\varepsilon_L \dpod{R}\zeta_L\,.
\]
It is clear that the transformations involving $\varepsilon_R$ do not preserve the form of $\mathcal {B}$. 

In the expression (\ref{eq:Bfield}) the first term is the kinetic term for the left-handed field $\Bfield$ \eqref{eq:Bfield}, $\gamma$ is the real-valued parameter of the deformation, and the latter term has explicit dependence on the Lagrange multiplier field $\const$. Combined together with \eqref{eq:LagrNonChiral} the new constraint on the isovector superfield reads
\[\label{eq:ConstrSuperfield}
\vecs^i\vecs_i = r _ 0 + 2 \g \mathcal {B},
\]
which only changes the latter two constraints from \eqref{eq:SphereConstr}, namely
\<
\psi_L^i n ^ i \eq \gamma \zeta_L\,, \nln\
F^i n^i - i\psi^i_L \psi^i_R\eq \g G\,,
\>
leaving the first constraint and the constraint for the right-handed component of $\psi _ R$ intact. The full Lagrangian is given by the expression
\bea
\mathcal {L} & = &  2 \p _ L n^ i \p _ R n^ i + i \psi _ L ^ i \p _ R \psi _ L ^ i
+ i \psi _ R ^ i \p _ L \psi _ R ^ i + \half F ^ {i 2}  \nonumber \\
& + & i \z _ L \p _ R \z _ L + \half G ^ 2+ i \gamma \lambda_R \zeta_L + \gamma G\sigma \nonumber \\
& - & \s \l ( F ^ i n^ i - i \psi ^ i _ L \psi ^ i _ R \r ) 
- \half D \l ( n^ i n^ i - r _ 0 \r ) 
- i \lambda _ R \psi _ L ^ i n^ i + i \lambda _ L \psi ^ i _ R n^ i \,.
\eea
Integration over the auxiliary fields $F^i$ and $G$ yields
\<\label{eq:LagrDynamic}
\lagr \eq 2\p_L n^i \p_Rn^i + i \psi _ L ^ i \p _ R \psi _ L ^ i+ i \psi _ R ^ i \p _ L \psi_R^i \nl
+ i \z _ L \p _ R \z _ L  -i\lambda_R(\psi^i_Ln_i-\gamma\zeta_L)  + i \lambda _ L \psi ^ i _ R n^ i \nl
-\half\gamma^2\sigma^2 - \half \left(D + \sigma ^ 2 \right ) n ^ i n ^ i +\half D r _ 0 + i\sigma\psi^i_L\psi^i_R\,.
\>
%

\section{Effective Potential and Vacua}\label{Sec:EffPotVac}

Given \eqref{eq:LagrDynamic} we first wish to find the vacua of the theory similarly to \cite{Witten:1978bc} where the nonsupersymmetric $U(N)$ sigma model was solved in the large-$N$ limit.  Integrating out $n^i$ and $\psi^i$ fields and expressing the result in terms of the dynamical scale $\Lambda$ related to the bare coupling constant by
\be
r _0 = \f {N} {4 \pi} \log \f {M _ {UV} ^ 2} {\Lambda ^ 2},
\ee
with $M _ {UV}$ being the ultraviolet cutoff, we obtain the following effective potential provided the rest of the fermionic fields are put to zero
\[\label{eq:EffectivePotential}
V_{eff} = \frac{N}{8\pi}\left[D\log\frac{\Lambda^2}{D+\sigma^2}+\sigma^2\log\frac{\sigma^2}{\sigma^2+D}+D+u\sigma^2\right]\,,
\]
where we have introduced a new deformation parameter
\[
u=\frac{4\pi\gamma^2}{N}\,.
\]
Minimizing the potential with respect to $D$ and $\sigma$ one finds the vacua of the theory
\<\label{eq:VacuumSolution}
\sigma _ 0 \eq \pm\Lambda \mathrm{e}^{-\sfrac{u}{2}}\,, \nln
D \eq \Lambda^2 - \sigma^2\,.
\>
Thus, there are two different vacua in the deformed model for any finite parameter $u$. One can integrate out the $D$ field \eqref{eq:EffectivePotential} and obtain the potential which depends only on $\sigma$
\[\label{eq:EffectivePotentialsigma}
 V_{eff}= \frac{N}{8\pi}\left[\Lambda^2+\sigma^2\left(\log\frac{\sigma^2}{\Lambda^2}-1+u\right)\right]\,.
\]
%

\paragraph{Vacuum Energy.}

Plugging the vacuum solution \eqref{eq:VacuumSolution} into \eqref{eq:EffectivePotential} we calculate the vacuum energy
\[
\energy_{vac} = \frac{N}{8\pi}\Lambda^2\left(1-\mathrm{e}^{-u}\right)\,.
\]
From the expression above it is obvious that the supersymmetry is unbroken only if $u=0$. At small $u$ the vacuum energy behaves linearly with  $u$ or quadratically with the deformation parameter $\gamma$
\[\label{eq:VacuumEnergy}
 \energy_{vac} \sim \frac{N}{8\pi}u\Lambda^2 = \frac{\gamma^2}{2}\Lambda^2\,.
\]
At large $u$ the vacuum energy scales linearly with $N$
\[
 \energy_{vac} \sim \frac{N}{8\pi}\Lambda^2 \,.
\]
%

\section{Spectrum}\label{Sec:Spectrum}

\begin{figure}
\begin{center}
\includegraphics[height=2cm, width=4cm]{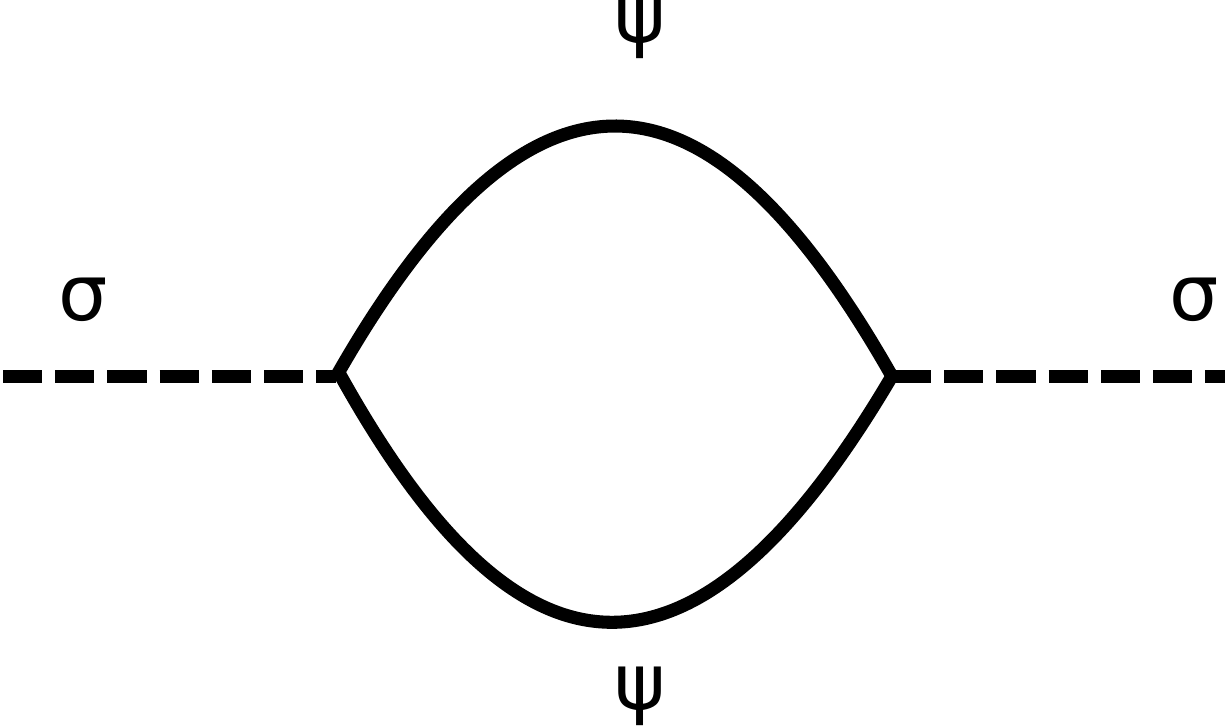}
\quad \includegraphics[height=2cm, width=4cm]{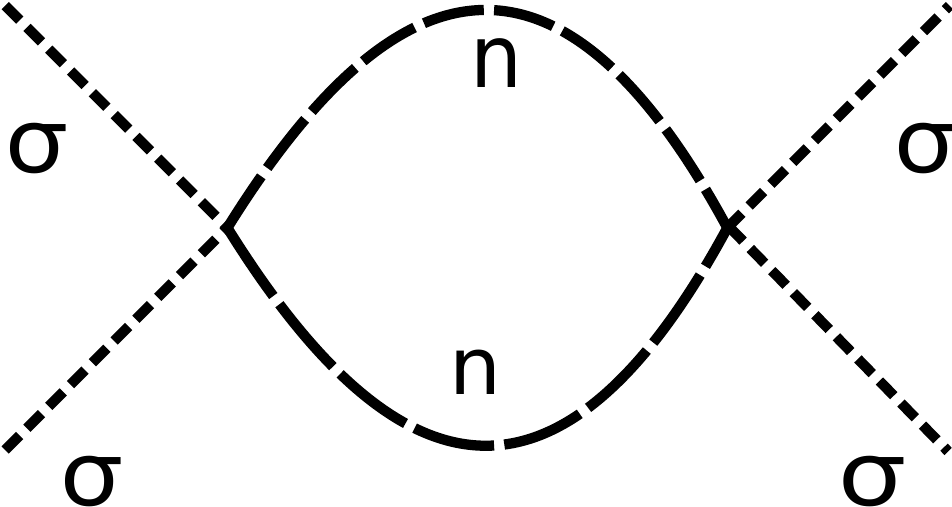}
\caption{Feynman diagrams contributing to the wave function renormalization of $\sigma$.}
\label{fig:sigmawaveren}
\end{center}
\end{figure}

\begin{figure}
\begin{center}
\includegraphics[height=2cm, width=4cm]{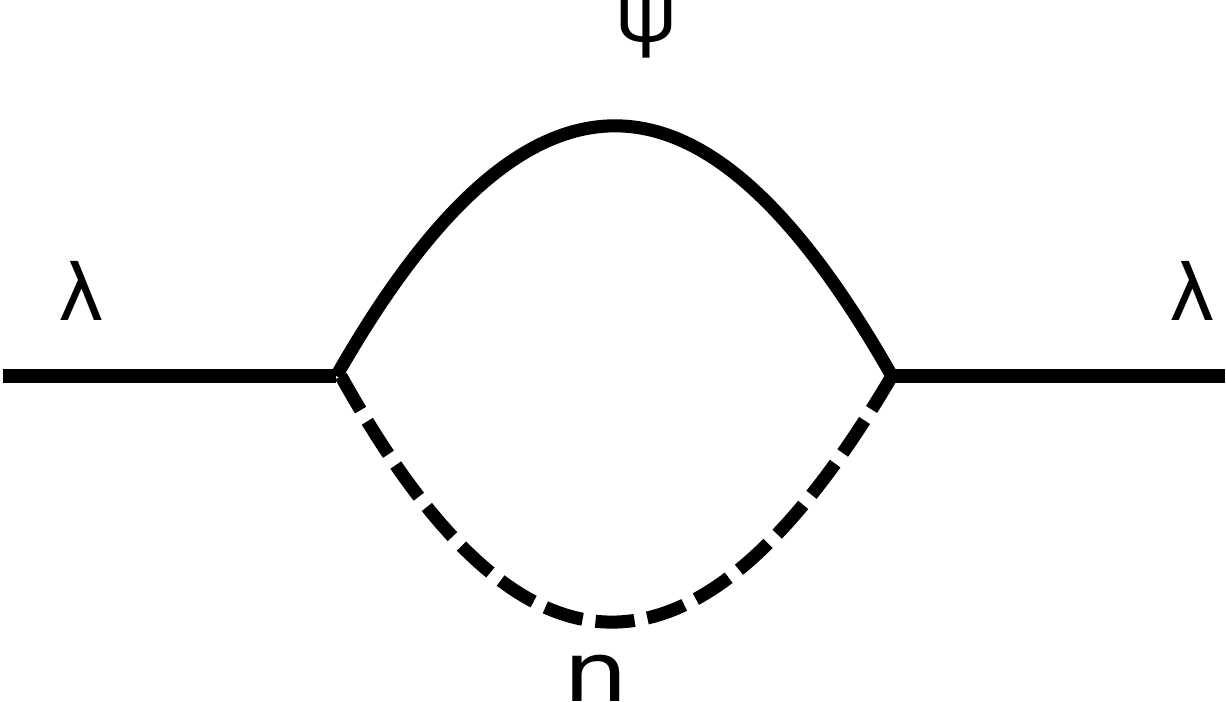}
\caption{The wave function renormalization for $\lambda$.}
\label{fig:lambdawaveren}
\end{center}
\end{figure}

\begin{figure}
\begin{center}
\includegraphics[height=3cm, width=3cm]{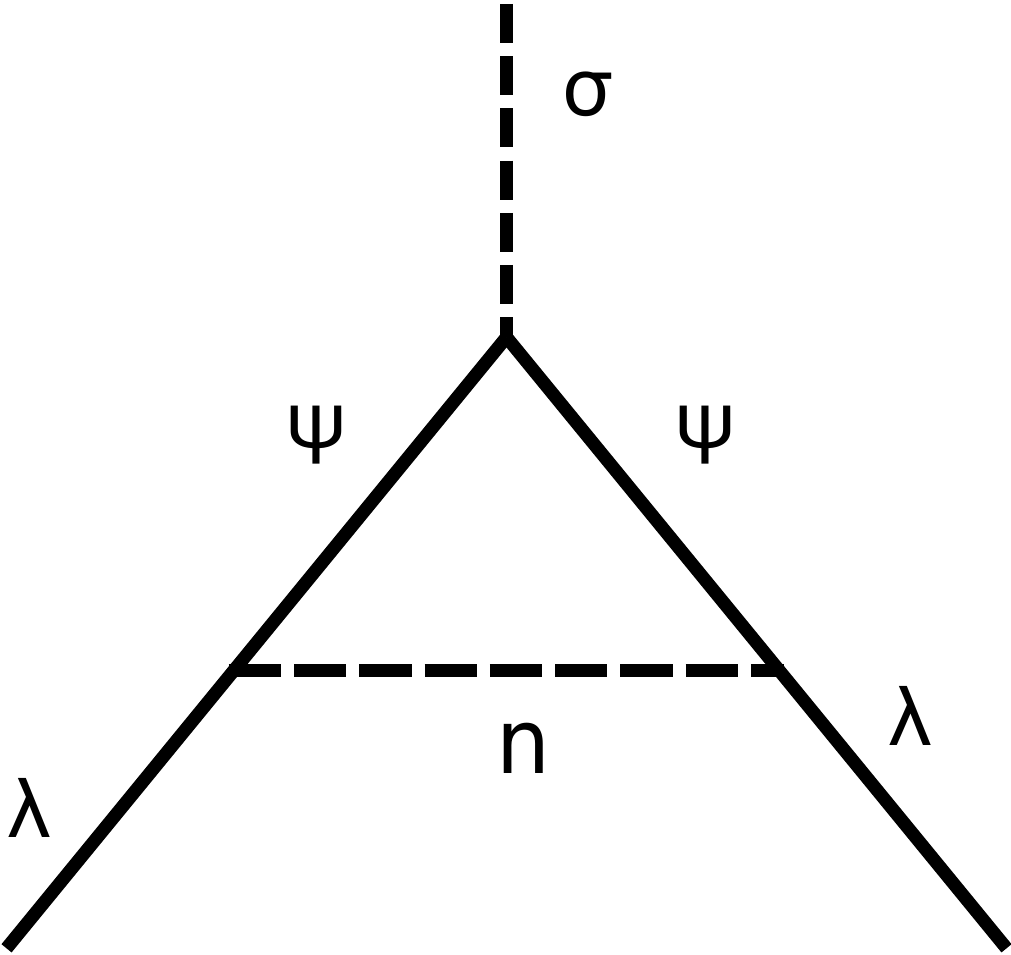}
\caption{The induced Yukawa vertex.}
\label{fig:yukawa}
\end{center}
\end{figure}

The next goal is to find the spectrum of the theory. To do so we are to obtain the one-loop effective action. The most straightforward way to calculate the action is to consider small fluctuations of the fields around the vacuum and to use what is called the long wave approximation. As a result we get the one-loop effective action in the large-$N$ approximation for the fields $\s$, $\lambda$ and $\zeta$
\bea
\mathcal {L} _ {eff} =  \f {1} {2 e _ \s ^ 2} \l ( \p _ \m \s \r ) ^ 2 + \f {i} {2 e _ \lambda ^ 2} \bar \lambda \g ^ \m \p _ \m \lambda - 
V _ {eff} (\s) + i \z _ L \p _ R \z _ L + \f {1} {2} \Gamma \s \bar \lambda \lambda + i \g \lambda _ R \z _ L,
\label{efflag}
\eea
where $e _ \s$ and $e _ \lambda$ are the coupling constants that define the wave function renormalization of the $\s$ and $\lambda$ fields correspondingly, and $\Gamma$ is induced the Yukawa coupling of the $\lambda$ and $\s$. The wave function renormalization is easily calculated in the limit of small momenta. The diagrams contributing to $e _ \s$ are shown in \figref{fig:sigmawaveren}
The actual calculation yields
\bea
\f {1} {e _ \s ^ 2} & = & \f {N} {8 \pi} \l ( \f {2} {3} \f {\s _ 0 ^ 2} {(\s _ 0 ^ 2 + D ) ^ 2} + 
\f {1} {3} \f {1} {\s _ 0 ^ 2} \r ) = 
\frac{N}{24\pi}\frac{\mathrm{e}^{u}}{\Lambda^2}\left[1 + 2 \mathrm{e}^{-2u}\right].
\label{eq:RenormSigma}
\eea
Similarly, the renormalization of $\lambda$ is given by the diagram in \figref{fig:lambdawaveren}.
\bea
\f {1} {e _ \lambda ^ 2} & = & \f {N} {4 \pi} \l ( \f {1} {D} - \f {\s _ 0 ^ 2} {D ^ 2} 
\log \f {\s _ 0 ^ 2 + D } {\s _ 0 ^ 2} \r ) = \frac{N}{4\pi}\frac{1}{\Lambda^2}\frac{1-\mathrm{e}^{-u}(1+u)}{(1-\mathrm{e}^{-u})^2}.
\label{eq:Renormlambda}
\eea
Finally, the Yukawa coupling $\Gamma$ can be found from either the triangular graph (see \figref{fig:yukawa}), or, equivalently, as the masses renormalization from the diagram in \figref{fig:lambdawaveren},  
\bea
\Gamma & = & \f {N} {4 \pi} \f {1} {D} \log \f {\s _ 0 ^ 2 + D } {\s _ 0 ^ 2} = \frac{N}{4\pi}\frac{1}{\Lambda^2}\frac{u}{1-\mathrm{e}^{-u}}.
\label{eq:RenormGamma}
\eea
It is worth noting that although the fields $\s$ and $\lambda$ were introduced as auxiliary dummy fields in \eqref{eq:HeteroticDeformation}, they become dynamical after integrating out the fields $n$ and $\psi$. It is clear that in the limit $D \to 0$ or $u \to 0$, which corresponds to the restoration of the supersymmetry, the coupling constants $e _ \s$ and $e _ \lambda$ coincide
\be
\f {1} {e _ \s ^ 2} = \f {1} {e _ \lambda ^ 2} = \f {N} {8 \pi} \f {1} {\s _ 0 ^ 2}.
\ee
Now we can turn to actual calculation of the mass spectrum. It is obvious that the $n$ and $\psi$ fields acquire mass due to the VEV of $D$ and $\s$, namely
\bea
m_ \psi ^ 2 & = & \s _ 0 ^ 2, \nonumber \\
m _ n ^ 2 & = & \s _ 0 ^ 2 + D.
\eea
From the effective Lagrangian (\ref{efflag}) one can easily find the expressions for the masses of the auxiliary field $\sigma$
\bea
m _ \s = \Lambda \sqrt {6} \f {e ^ {u / 2}} {\sqrt {1 + \half e ^ {2 u}} }.
\eea
At nonzero values of the heterotic deformation parameter mixing between $\zeta_L, \lambda_L$ and $\lambda_R$ occurs. In order to find the mass states, one needs to transform the fermion mass matrix to the canonical form. However, it is clear that there is a massless mode since only the left fermion was introduced. To see this, one has to find the solution of the characteristic polynomial, corresponding to the fermion mass matrix
\be
m \l ( m ^ 2 - \g ^ 2 e _ \lambda ^ 2 - \s _ 0 ^ 2  e _ \lambda ^ 4 \Gamma ^ 2 \r ) = 0,
\ee
which indeed has zero solution for any $u$. There is also another solution of the characteristic polynomial. Summarizing, we find the following masses of the fermions 
\bea
m _ F & = & 0, \nonumber \\
m _ F & = & \sqrt {\g ^ 2 e _ \lambda ^ 2 + \s _ 0 ^ 2  e _ \lambda ^ 4 \Gamma ^ 2  } = 
2 \Lambda \f {\sqrt {u \l ( e ^ u - 1  \r ) } } {e ^ u - 1 - u} \, \sinh \f {u} {2}.
\label{fmas}
\eea

\subsection{Spectrum of masses at small $u$}

First, when the supersymmetry is unbroken $u=0$, $D=0$ the masses of superpartners coincide. For the fields $n^i$ and $\psi^i$ it become 
\[
m_{\psi_{L,R}}=m_n=\Lambda,
\]
while for the fermion $\lambda_{L,R}$ and boson $\sigma$ we have 
\[
m_{\lambda_{L,R}}=m_\sigma=2\Lambda\,.
\]
The $\zeta_L$ field is decoupled from other fields, it is sterile and massless.

\subsection{Spectrum of masses at large $u$}

\begin{figure}
\begin{center}
\includegraphics[height=4cm, width=4cm]{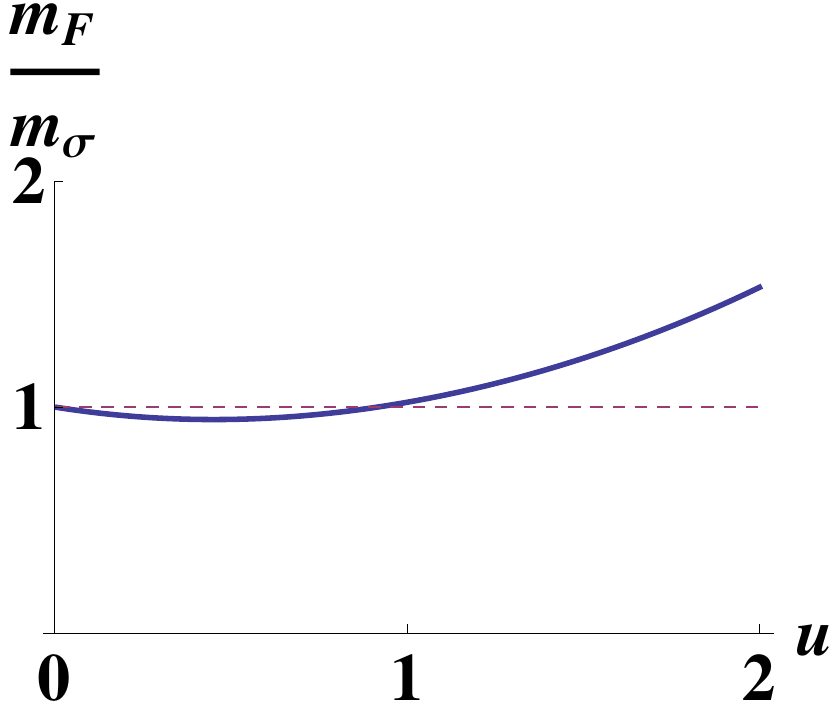}
\caption{The ratio of the fermion mass matrix eigenvalue to the mass of $\sigma$ as a function deformation parameter $u$.}
\label{fig:ratio}
\end{center}
\end{figure}

When the parameter of heterotic deformation gets bigger, the splitting between the masses in our theory becomes more dramatic. For the $n^i$ and $\psi^i$ particles they are
\[
m_n= \sqrt{D+\sigma^2} = \Lambda\,,\quad m_\psi = \Lambda\mathrm{e}^{-\sfrac{u}{2}}\,,
\]
thus the fermions become much lighter than bosons. The couplings behave differently as well. The coupling for $\sigma$ is
\[
\frac{1}{g_\sigma^2}=\frac{N}{4\pi}\frac{\mathrm{e}^u}{6\Lambda^2}\,,
\]
and for $\lambda$ it becomes
\[
\frac{1}{g_\lambda^2} = \frac{N}{4\pi}\frac{1}{\Lambda^2}\,.
\]
The mass of the $\s$ field goes exponentially to zero for large $u$
\be
m _ \s = 2 \sqrt {3} \Lambda e ^ {-u/2}.
\ee
The fermion mass matrix has a zero eigenvalue corresponding to the now massless left component of the field $\lambda _ L$, while the mixture of the fields $\z _ L$ and $\lambda _ R$ produce the mass term
\be
m _ { \lambda _ R, \z _ L } = \Lambda \sqrt {u}.
\ee  
We see that being equal in the limit $u \to 0$ the masses of $\s$ and $\lambda$ now become essentially different. For arbitrary value of the deformation parameter the ratio of the fermion matrix eigenvalue and the mass of $\s$ is plotted in \figref{fig:ratio}.
Although the masses become equal not only for $u=0$, there is no restoration of the supersymmetry. First, there is still a Goldstino and second, the vacuum energy (\ref{eq:VacuumEnergy}) is not equal to zero at that point.

\section{Superfield formulation of the heterotic $\mathcal{N}=(0,2)$ $\mathbb{CP}^{N-1}$ sigma model}\label{Sec:SuperfieldCPN}


Distler and Kachru \cite{Distler:1993mk} constructed a very wide class of heterotic $(0,2)$ Calabi-Yau sigma models and provided gauged formulation for them. Their construction employs fiber bundles over complex projective spaces and includes heterotic $(0,2)$ $\mathbb{CP}^{N-1}$ sigma model in it. In this section we give a simple alternative derivation of the $(0,2)$ heterotic (weighted) $\mathbb{CP}^N$ sigma model from the Majorana formalism we used above.  

Recall that for the $\text{O}(N)$ sigma model we used the following trick -- the constraint on the isovector superfield was replaced by a different one, but it did not affect the lowest component constraint; thereby the geometry of the theory was not deformed. In order to apply this trick for the $\mathbb{CP}^{N-1}$ model we write its action in the form found in \cite{D'Adda:1978kp, Krivoshchekov:1984qm} and deform it by adding the coupling of chiral field 
\be
\Bfield = - \bar \t \z + \half \bar \t \t \bar {\mathcal {F}} \mathcal {F},
\ee
constraint superfield 
\be
\mathcal {S} = \sqrt {2} \s _ 1 + \sqrt {2} \bar \t u + \half \bar \t \t D
\ee
and spinor superfield
\[
\mathcal{A}_\alpha = -i (\gamma^\mu \theta)_\alpha A_\mu + \sqrt {2} (\gamma^5\theta)_\alpha \s _ 2 
+ \sqrt {2} \bar{\theta}\theta\, v _ \a\nln,
\label{gauge_field}
\]
where $\t$ is a Majorana spinor, $\s _ 1$, $\s _ 2$, $A _ \m$, $u _ \a$, $v _ \a$ and $D$ are real fields, while $\z$ and $\mathcal{F}$ are complex ones \footnote{Note that for the present section we changed the notations of the fields in order for the reader to see the equivalence with the Lagrangian from \cite{Shifman:2008kj} more easily. Also for convenience we consider $\z$ to be a right-handed fermion and use the definition 
$\p _ {L,R} = \p _ 0 \pm \p _ 1$ instead of one mentioned in Appendix.}. Introducing now the complex isovector superfield
\[
\vecs^i = n^i + \bar{\theta}\xi ^ i + \half\bar{\theta}\theta F^i,
\]
we can write the Lagrangian of the model in the following form 
\<\label{eq:N1Heterotic}
\lagr_{\mathbb{CP}^N} \eq \int d^2 \theta\, \Big[\sfrac{1}{2}\varepsilon_{\beta\alpha}(\covder_\alpha+i\mathcal{A}_\alpha){\vecs}^\dag_ i (\covder_\beta-i\mathcal{A}_\beta)\vecs_i + i \const(\vecs_i^\dag \vecs_i - r _ 0)  \nl 
 + \sfrac{1}{4}\varepsilon_{\beta \alpha}\covder_\alpha\Bfield^\dag\,\covder_\beta\Bfield
+\l ( i \, \omega \, \Bfield (\const-\sfrac{i}{2}\overline{\covder}\gamma^5\mathcal{A}) + \text{H.c.} \r ) \Big ]\,,
\>
where $\omega$ is the complex-valued deformation parameter and 
$\overline{\covder}\gamma^5\mathcal{A} = \covder _ \a \l ( \g ^ 0 \gamma^5 \r ) _ {\a \b} \mathcal{A} _ \b$. Some comments about the Lagrangian are due. The advantage of the superfield formulation is that the supersymmetry is manifest without an explicit check. Although the Lagrangian for the undeformed theory is written using the $\mathcal{N} = (1,1)$ formulation it possesses $\mathcal{N} = (2,2)$ symmetry due to the K\"{a}hler structure of the target space \cite{Zumino:1979et}. The field $\mathcal{B}$ is a superfield only with respect to the half of supertransformations $\t _ R \to \t _ R + \varepsilon _ R$, therefore the symmetry of the deformed Lagrangian is $\mathcal {N} = (0,2)$ one. Also it should be noted that the field $\mathcal {A} _ \a$ has the form (\ref{gauge_field}) only if one considers a particular gauge. Starting from the most general expression for the real spinor field
\be
\mathcal{A}_\alpha = a _ \a + A \t _ \a - i (\gamma^\mu \theta)_\alpha A_\mu + \sqrt {2} (\gamma^5\theta)_\alpha \s _ 2 
+ \bar{\theta}\theta \l( v _ \a + \sfrac {i} {2} (\g ^ \m \p _ \m a) _ \a \r ),
\ee 
one can use the following gauge transformations
\[
\mathcal{A}_\alpha \to \mathcal{A}_\alpha - \covder_\alpha \Phi,
\]
with $\Phi$ being the scalar real superfield, to eliminate $a _ \a$ and $A$. The undeformed Lagrangian is obviously gauge invariant, while the invariance of the deformation is more subtle. The transformation of the term $\covder _ \a \l ( \g ^ 0 \gamma^5 \r ) _ {\a \b} \mathcal{A} _ \b$ is proportional to
$\l ( \g ^ 0 \gamma^5 \r ) _ {\a \b} \covder _ \a \covder _ \b \Phi$, which is identically zero since the operators $\covder _ 1$ and $\covder _ 2$ anticommute.

Carrying out the calculations and integrating out the auxiliary fields $F$ and $\mathcal {F}$ we recover the Lagrangian of the $(0,2)$ $\mathbb{CP}^{N-1}$ sigma model considered in \cite{Shifman:2008kj}
\bea
\mathcal {L}_{\mathbb{CP}^N} & = & \l |\nabla _ \m  n _ i \r | ^ 2 + i \bar \xi _ L ^ i \nabla _ R \xi _ L ^ i
+ i \bar \xi _ R ^ i \nabla _ L \xi _ R ^ i - 2 | \s | ^ 2 |n _ i| ^ 2
- D \l ( |n _ i | ^ 2 - r _ 0 \r ) \nonumber \\ 
& + & \l [ i \sqrt {2} \bar n _ i \l ( \lambda _ L \xi _ R ^ i- \lambda _ R \xi _ L ^ i \r )  - i \sqrt {2} \s \bar \xi _ R ^ i \xi _ L ^ i + \text {H.c.} \r ] - 4 |\omega| ^ 2 |\s| ^ 2 \nonumber \\
& + & \sfrac{i} {2} \bar \z _ R \p _ L \z _ R - \l [ i  \sqrt {2} \omega \lambda _ L \z _ R + \text {H.c.} \r ],
\eea
where the following complex fields have been introduced
\[
\sigma = \s _ 1 + i \s _ 2 \,, \quad \lambda _ \a = u _ \a + i v _ \a,
\]
and $\nabla _ \m = \p _ \m - i A _ \m$ is a usual notation for the covariant derivative.

Analogously, the weighted $\mathbb{CP}^N$  sigma-model considered in \cite{Shifman:2010id} which emerges from the reduction of $\mathcal{N} = 1$ supersymmetric QCD with gauge group $\text {U} (N _ c)$ and $N _ f$ flavors can be easily deformed to the chiral version by \eqref{eq:N1Heterotic}, where the constraint $\vecs_i^\dag \vecs_i=r _ 0$ is replaced by 
\[
\sum\limits_{i=1}^ {N _ c} \vecs_i^\dag \vecs_i-\sum\limits_{i=N _ c +1}^{N _ f} \vecs_i^\dag \vecs_i=r _ 0~.
\]
In components the Lagrangian reads
\bea
\mathcal {L}^{\text{w}}_{\mathbb{CP}^N} & = & \l |\nabla _ \m  n _ i \r | ^ 2 + \l |\nabla _ \m  \rho _ i \r | ^ 2 + i \bar \xi _ L ^ i \nabla _ R \xi _ L ^ i
+ i \bar \xi _ R ^ i \nabla _ L \xi _ R ^ i + i \bar \eta _ L ^ i \nabla _ R \eta _ L ^ i
+ i \bar \eta _ R ^ i \nabla _ L \eta _ R ^ i\nonumber \\ 
& - & 2 | \s | ^ 2 |n _ i| ^ 2 -2|\sigma|^2 |\rho_i| ^ 2
- D \l ( |n _ i | ^ 2 - |\rho _ i| ^ 2 -  r _ 0 \r ) - 4 |\omega| ^ 2 |\s| ^ 2 \nonumber \\ 
& + & \l [ i \sqrt {2} \bar n _ i \l ( \lambda _ L \xi _ R ^ i- \lambda _ R \xi _ L ^ i \r )  - i \sqrt {2} \s \bar \xi _ R ^ i \xi _ L ^ i + \text {H.c.} \r ]  \nonumber \\
& + & \l [ - i \sqrt {2} \bar \rho _ i \l ( \bar \lambda _ L \eta _ R ^ i - \bar \lambda _ R \eta _ L ^ i \r )  + i \sqrt {2} \bar \s \bar \eta _ R ^ i \eta _ L ^ i + \text {H.c.} \r ]  \nonumber \\
& + & \sfrac{i} {2} \bar \z _ R \p _ L \z _ R - \l [ i  \sqrt {2} \omega \lambda _ L \z _ R + \text {H.c.} \r ],
\eea
with
\bea
\vecs^i & = & n^i + \bar{\theta}\xi ^ i + \half\bar{\theta}\theta F^i, \quad i=1,\dots,N _ c, \nonumber \\
\vecs^{ N _ c + i } & = & \rho^i + \bar{\theta}\eta ^ i + \half\bar{\theta}\theta F^i, \quad i = 1,\dots, N _ f - N _ c.
\eea


\section{Conclusions}\label{Sec:Conclusions}

In this paper we have constructed the heterotic $\mathcal {N} = (0,1)$ supersymmetric $\text{O}(N)$ sigma model. Similarly to the $\mathbb{CP}^N$ sigma model it can be solved at large $N$. As a result we found the effective potential of the theory, which allowed us to get the spectrum. It appeared to be very much reminiscent of the spectrum of the $(0,2)$ $\mathbb{CP}^N$ sigma model in spite of the fact that the latter possesses a K\"{a}hler structure, and therefore a larger supersymmetry, whereas the former does not. For all values of the deformation parameter there is a massless fermion. For $\gamma = 0$ it is the extra left-handed sterile fermion, while for $\g \neq 0$ it is a Goldstino (mixture of the additional left-handed fermion and the initial right-handed one) corresponding to the supersymmetry breaking. The existence of the Goldstino is the evidence of the fact that the supersymmetry becomes broken at any nonzero value of the deformation parameter. Another proof of the supersymmetry breaking is the presence of the nonzero vacuum energy density. Also fields from the same multiplet acquire different masses when the deformation is turned on.

The low energy one-loop effective potential has two vacua. From the naive point of view there should be one kink (and one antikink) interpolating between those vacua. However, such an argument leads to a wrong conclusion about the number of kinks for the $\mathbb{CP}^N$ sigma model. For the latter kink dynamics has been studied to a very high extent (see \cite{Hanany:2004ea,Dorey:1998yh, Hori:2000kt} and \cite{Shifman:2010id}). It was found that the different kinks correspond to the different integration contours in the $\s$ plane and that the number of kinks interpolating between two different vacua depends not only on $N$ but also on the number of vacua separating those two. Therefore, the question about the number of kinks in the $\text{O}(N)$ theory should be worked out more carefully.    

In the current paper we used Majorana $\mathcal{N}=1$ superfield formalism while constructing the action of the $\text{O}(N)$ sigma-model. The heterotic deformation was rendered by the coupling of the extra chiral superfield to the auxiliary superfield we have used to build up the undeformed theory. We then generalize this construction to the heterotic $(0,2)$ $\mathbb{CP}^N$ and weighted $\mathbb{CP}^N$  sigma models by adding an additional auxiliary superfield and modifying the interaction between the chiral superfield and the auxiliary superfields.  One may now try to use our methods for investigating sigma models with twisted masses and their chiral deformations.

Perhaps at this point the most intriguing question to be answered is from which four-dimensional bulk theory does the $\text{O}(N)$ sigma model originate from (if any). The analogy with $\text{U}(N)$ Yang-Mills theory and $\mathbb{CP}^{N-1}$ sigma models discussed in \cite{Gorsky:2004ad} and references therein is not clear -- what number of supersymmetries does the bulk theory have to have? It is interesting to understand by means of what (extended object or mechanism) the bulk supersymmetry gets broken.

Finally, the results of this paper hopefully can be used in string theory applications.

\section*{Acknowledgments}

We would like to thank M. Shifman for pointing out the existed problem. We are also indebted to  M. Voloshin, A. Vainshtein and A. Yung for illuminating and elucidating discussions. This work was supported in part by the DOE grant DE-FG02-94ER40823 (P.K.); by the Stanwood Johnston grant from the Graduate
School of University of Minnesota, RFBR Grant No. 07-02-00878, and by the
Scientific School Grant No. NSh-3036.2008.2. (A.M.).

\appendix

\section{Notations}\label{Sec:Notations}

Here we list the notations and some useful relations we use in the paper.

Gamma matrices
\[
\gamma^0 =  \sigma_2 =
\begin{pmatrix}
0& -i\\
i&0
\end{pmatrix}\,, \quad
\gamma^1 =  i\sigma_1 =
\begin{pmatrix}
0& i\\
i&0
\end{pmatrix}\,,\quad
\gamma^5 = \gamma^0 \gamma^1 = \sigma_3 =
\begin{pmatrix}
1& 0\\
0&-1
\end{pmatrix}\,.
\]
Antisymmetric symbol
\be
\varepsilon _ {\a \b} = \l (
\ba{cc}
0 & 1 \\
-1 & 0 
\ea
\r ).
\ee
Left and right coordinates
\[
\begin{array}[b]{rclcrclcrcl}
x_L \eq x_0 + x_1, &&
\p _ 0 \eq \p _ L + \p _ R, &&
\p _ L \eq \half \l ( \p _ 0 + \p _ 1 \r ),
\\
x_R \eq x_0 - x_1, &&
\p _ 1 \eq \p _ L - \p _ R, && 
\p _ R \eq \half \l ( \p _ 0 - \p _ 1 \r )\,.
\\
\end{array}
\]
Left and right fermions
\<
\psi \eq 
\l(
\ba{c}
\psi _ R\\
\psi _ L
\ea
\r )
\>
are eigenstates of $\gamma^5$
\[
\gamma^ 5 \psi _ {R,L} = \pm \psi _ {R,L}\,.
\]
Derivatives and integrals
\<
\int d^ 2 \t \,\bar \t \t \eq \int d \t _ 1 \, d \t _ 2 \bar \t \t = \int d \t _ 1 \, d \t _ 2 \, 2 i \t _ 2 \t _ 1 = 2 i\,, \nln
\f { \p } {\p \bar \t _ \a } \t _ \b \eq \g ^ 0 _ {\a \b}\,.
\>
Contraction of indices for Majorana fermions
\be
\bar { \psi } \t = \psi ^ \dagger \g ^ 0 \psi = \psi ^ T \g ^ 0 \psi 
=i \t _ 2 \psi _1 - i \t _ 1 \psi _2 = \bar {\theta} \psi\,,
\ee
\be
\bar {\t} \g ^ {0,1} \t = \l ( \t _ 1 \r ) ^ 2 = \l ( \t _ 2 \r )^ 2 = 0\,,
\ee
\bea
\bar {\t}  \t = 2 i \t _ 2 \t _ 1 = - 2 i \t _ 1 \t _ 2\,, \nonumber \\
\t _ \a \t _ \b = \f {i} {2} \epsilon _ {\a \b} \, \bar {\t}  \t = - \half
\g ^ 0 _ {\a \b} \bar {\t}  \t\,, \nonumber \\
\bar {\t} _ \a  \t _ \b = \half \delta _ {\a \b} \, \bar {\t}  \t\,.
\eea
Some relations for gamma matrices
\<
\g^{\m T} \eq - \g ^ 0 \g ^ \m \g ^ 0\,, \nln
\g^{\m \dagger} \eq \g ^ 0 \g ^ \m \g^0\,.
\>
%

\paragraph{Supersymmetry transformations.}

Coordinate transformations 
\bea
x _ \m & \to & x _ \m + i \bar \eps \g  _ \m \t, \nonumber \\
\t _ \a & \to & \t _ \a + \eps _ \a, \nonumber \\
\bar {\t} _ \a & \to & \bar { \t } _ \a + \bar { \eps } _ \a.
\eea
Chiral superfield 
\[\label{eq:ChiralSuperField}
\Phi = \phi+ \bar{\theta}\psi + \half \bar{\theta}\theta F\,, 
\]
obeys the following supertransformations
\<
\delta n\eq \bar {\eps} \psi, \nln
\delta \psi \eq -i \p _ \m n\g ^ \m \eps + \eps F, \nln
\delta F \eq - i \bar {\eps} \g ^ \m \p _ \m \psi\,.
\>
A natural generalization of the chiral superfield is the isovector superfield
\[
\vecs^i = n^i+ \bar{\theta}\psi^i + \half \bar{\theta}\theta F^i\,,\quad  i=1,\dots, N\,.
\]
Supertransformations act as
\[
\delta \Phi = \bar \epsilon \gen{Q} \Phi\,
\]
Supersymmetry generators
\<
\gen{Q}_\alpha \eq \f {\p} {\p \bar \t _ \a } -i \left(\gamma^\mu \theta \right)_\alpha \dpod{\mu} \,,
\>
Covariant derivative
\[
\covder_\alpha = \f { \p } {\p \bar \t _ \a } +i \l ( \g ^ \m \t \r ) _ \a \dpod{\mu}\,,
\]
anticommutes with the supercharge
\[
\{\gen{Q}_\alpha, \covder_\beta\} = 0\,.
\]
%

\paragraph{Chiral Notation.}

One can use the following identification
\[
x^\mu = \gamma^\mu_{\alpha\beta} x^{\alpha\beta}\,,\quad \mu=1,2\,, \quad \alpha,\beta =1,2\,.
\]
Having done so we can write
\[
\gen{Q}_{\alpha} =\epsilon_{\alpha\beta} \po{\theta_\beta}+\theta_\beta\dpod{\alpha\beta}\,.
\]
Accordingly we have
\<
\acomm{\gen{Q}_1}{\gen{Q}_1} \eq 2\dpod{12} = 2\left(i\po{t}-i\po{x}\right)=2(\ham+\moment)\,,\nln
\acomm{\gen{Q}_2}{\gen{Q}_2} \eq 2\dpod{21} = 2\left(i\po{t}+i\po{x}\right)=2(\ham-\moment)\,,\nln
\acomm{\gen{Q}_1}{\gen{Q}_2} \eq 0\,,
\>
where $\ham$ and $\moment$ are energy and momentum charges respectively. Covariant derivative reads
\[
\covder_{\alpha} =i\epsilon_{\alpha\beta} \po{\theta_\beta}-i\theta_\beta\dpod{\alpha\beta}\,.
\]

\bibliography{cpn}
\bibliographystyle{nb}

\end{document}